\documentclass[times,12,twocolumn]{article}
\pdfoutput=1
\usepackage[english]{babel}
\usepackage{amsmath}
\usepackage{algpseudocode}
\usepackage{graphicx}
\usepackage[colorlinks=true, allcolors=blue]{hyperref}
\usepackage{subcaption}
\usepackage{tikz}
\usepackage{enumitem}
\usepackage[font=small,labelfont=bf]{caption}
\usepackage{authblk}
\date{}
\providecommand{\keywords}[1]
{
  \small	
  \textbf{\textit{Keywords---}} #1
}

 \title{A novel method to compute the contact surface area between an organ and cancer tissue}

    \author[1]{Alessandra Bulanti}
    \author[2]{Alessandro Carfì}
    \author[1,3,4]{Paolo Traverso}
    \author[1,3,4]{Carlo Terrone}
    \author[2,4]{Fulvio Mastrogiovanni}

    \affil[1]{{Department of Surgical and Diagnostic Integrated Sciences (DISC), University of Genoa},
		{Italy}
    }
    \affil[2]{{Department of Informatics, Bioengineering, Robotics, and Systems Engineering (DIBRIS), University of Genoa},
		{Italy}
    }
    \affil[3]{{IRCCS Policlinico San Martino, Joint Research Lab on Interaction Technologies for Minimally Invasive and Open Surgery},
		{Genoa},
		{Italy}
    }
    \affil[4]{{IO Surgical Research Spin-off University of Genoa},
		{Italy}
    }

\begin{document}  

\twocolumn[
  \begin{@twocolumnfalse}
\maketitle 
    \begin{abstract}
        With ``contact surface area" (CSA) we refers to the area of contact between a tumor and an organ. This indicator has been identified as a predictive factor for surgical peri-operative parameters, particularly in the context of kidney cancer. However, state-of-the-art algorithms for computing the CSA rely on assumptions about the tumor shape and require manual human annotation. In this study, we introduce an innovative method that relies on 3D reconstructions of tumors and organs to provide an accurate and objective estimate of the CSA. Our approach consists of a segmentation protocol for reconstructing organs and tumors from Computed Tomography (CT) images and an algorithm leveraging the reconstructed meshes to compute the CSA. With the aim to contributing to the literature with replicable results, we provide an open-source implementation of our algorithm, along with an easy-to-use graphical user interface to support its adoption and widespread use. We evaluated the accuracy of our method using both a synthetic dataset and reconstructions of 87 real tumor-organ pairs.
    \end{abstract}
 \keywords{ 3D Segmentation, Computed Tomography , Computer Science , Graphical User Interface }\\
  \end{@twocolumnfalse}
]

\section{Introduction}
\label{sec:Introduction}
In the surgical domain, the concept of \textit{contact surface area} (CSA) is used to identify the region between a tumor and the surrounding unaffected organ. It was introduced by Leslie et al. \cite{1} in 2014 in the field of urology and since then it has been increasingly considered an important parameter. In fact, CSA plays a vital role in medical surgeries, particularly in cases of kidney tumors, where it is extensively employed as a predictive factor for peri-operative outcomes. Therefore, it is of utmost importance and interest to identify precise methods for computing it.

In order to compute the CSA, Leslie et al. multiply the total area of the tumor, approximated as a sphere, by the percentage of intraparenchymal components, which was automatically measured using a 3D shape reconstruction technique. Two years later, Hsieh et al. \cite{2} proposed a straightforward approach to estimate the CSA. This proposal was aimed at overcoming an error in the method by Leslie et al., who erroneously assumed a direct proportionality between CSA and intraparenchymal percentage. The technique introduced by Hsieh et al. for CSA computation preserves the assumption of tumor sphericity, and adopts the formula $2 \times \pi \times r \times d$ where $r$ is the maximum radius of the tumor, and $d$ is the maximum depth of the tumor's intrusion into the uninvolved parenchyma. According to this method proposed by Hsieh, these two quantities are extracted from visual inspection of the sagittal or coronal planes of DICOM images obtained from either Computed Tomography (CT) or Magnetic Resonance Imaging (MRI). Subsequent studies have utilized the formula and the method introduced by Hsieh et al., analyzing 2D images to identify the maximum radius and depth, aiming to assess the predictive capacity of CSA for peri-operative parameters \cite{3,7,8,9,10}.

More recently, due to advancements in medical imaging technique, which has been made wildly available in software tools for the 3D reconstruction of organs, surgeons have shifted away from visual image analysis methods and begun to measure the CSA through 3D reconstructions of both kidney and tumor. 
Takagi et al. \cite{4} and Bianchi et al. \cite{5} evaluated the contact area by manually outlining the boundary of the tumor on the 3D reconstructed kidney.
Meanwhile, Umemoto et al. \cite{6} utilized a technique based on 3D reconstruction that enable to simulate the tumor removal and to estimate the CSA accordingly.

However, these studies rely on human intervention to define the CSA on the 3D reconstruction of the organ or tumor, leading to subjective estimates.
In our work, we aim to introduce a novel technique for CSA estimation to achieve a better precision and objectivity.
This method utilizes 3D reconstruction of the organ and the tumor, and through the geometry of the two reconstructed parts, it computes the CSA. Unlike previous methods, the calculations are not based on assumptions about tumor shape and performed on 2D images but the real geometries of tumor and organ are leveraged.
Our work presents the segmentation protocol adopted for reconstructing the 3D models of kidneys and tumors from CT scans, provides an in-depth description of the algorithm, reports a quantitative evaluation on a synthetic dataset, and a qualitative one using data from 82 patients. The algorithm implementation and a simple graphical user interface are publicly available to increase the likelihood of reproduce our work and adopting our technique in common practice.

\begin{figure}
    \centering
    \includegraphics[width=\columnwidth]{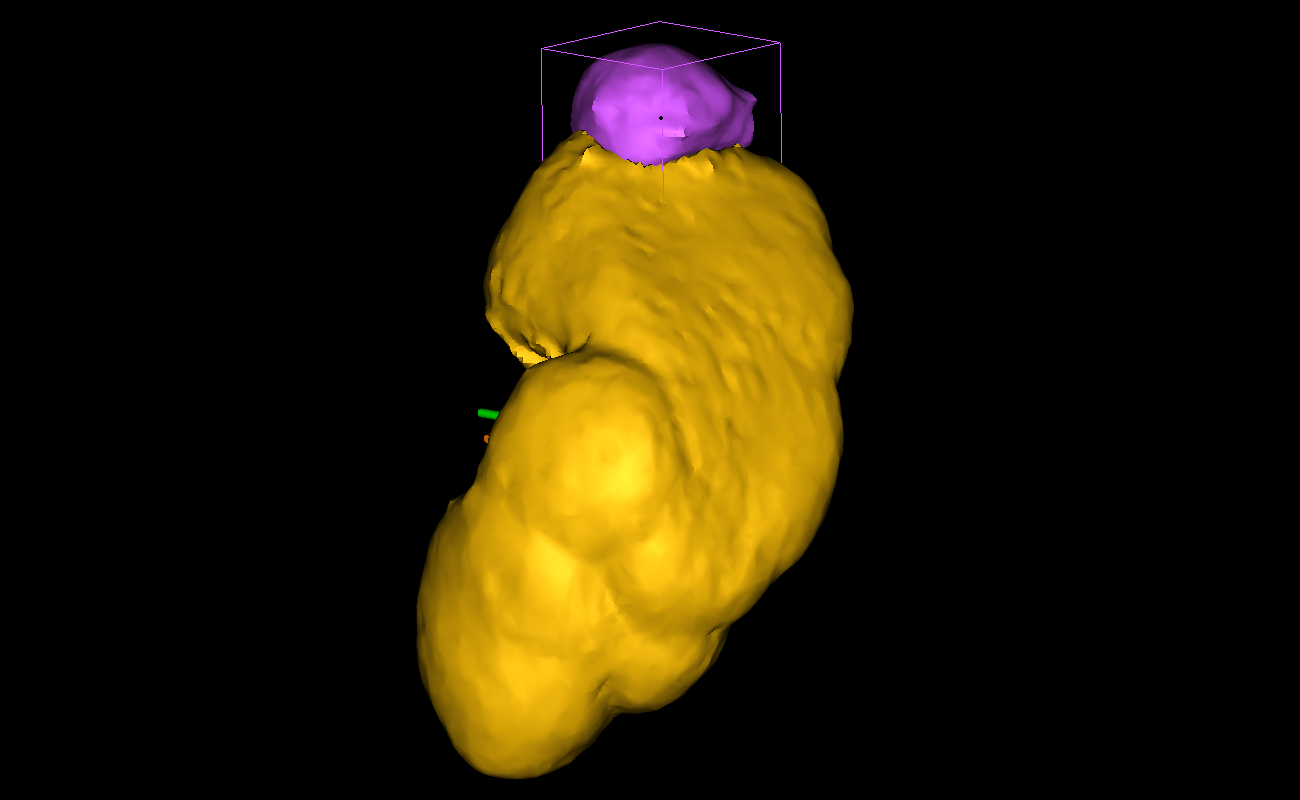}
    \caption{3D reconstruction of a kidney, in yellow, and a tumor, in purple.}
    \label{fig:finalreconstruction}
\end{figure}

\section{Segmentation Protocol}
\label{sec:Segmentation}
Our approach to computing the CSA relies on 3D models of the organ and the tumor. Therefore, the quality of the 3D reconstruction is crucial to ensure consistency in the CSA computation.
The quality of the reconstructions depends primarily on the accuracy of segmentation. Correctly scanning the region of interest and performing various operations on it with precision contribute to obtaining a more accurate reconstruction. Additionally, the resolution of the CT images. High-resolution images contain fewer artifacts, enabling more precise recognition of various regions of interest.

Currently, there is no fully automated segmentation available for the kidney, but researchers are working on it \cite{11,13,14}. 
Indeed this task is challenging due to the irregular structures of the kidney, which can vary significantly from person to person. Moreover, the presence of low contrast in some cases can introduce artifacts, hindering the precise reconstruction of the kidney structures. Finally, it is difficult to precisely identify kidney contours because scattered into various layers in the tomographic images\cite{11,12}.

Here, we outline the segmentation protocol for obtaining accurate 3D reconstructions of organs and tumors.
Our protocol is specific for kidneys and the associated tumor reconstructions.
The main objective of the protocol is to describe how the human operator should proceed to segment the kidney and the tumor to ensure an accurate reconstruction.
Therefore, we structure the protocol into three steps: kidney segmentation, tumor segmentation and reconstruction refinement. After completing these steps, accurate 3D reconstructions are generated (see Figure \ref{fig:finalreconstruction}) and exported as STL files. These files are then provided as input to our algorithm for calculating the CSA.

Our 3D reconstructions use CT DICOM images, which enable the visualization of the human body in three different planes: axial, coronal, and sagittal. It is noteworthy that in our implementation we adopt a Materialise Mimics InPrint\footnote{\url{https://www.materialise.com/en/healthcare/mimics-inprint}} from Materialise NV, but such a choice does not impose any contingent limitation to our segmentation protocol, which can in fact be applied in other scenarios and based on other medical-grade software applications.
\\

\begin{figure}[t]
    \centering
    \includegraphics[width=\columnwidth]{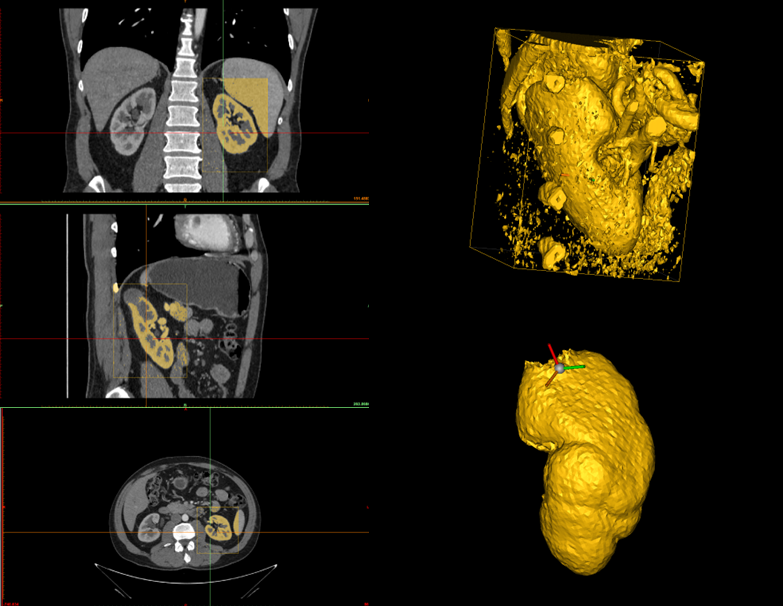}
    \caption{On the left hand side, the selection of the region of interest in the three planes(axial, coronal and sagittal) on the bases of the application of a threshold. On the right hand side, top, the 3D reconstruction after the threshold application. On the right hand side, bottom, the 3D reconstruction after manual cleanup.}
    \label{fig:kidneysegmentation}
\end{figure}

\textit{Kidney Segmentation}
\\This step requires both automated and manual operations from the operator.
First, the operator must navigate the CT scan to identify the kidney. Then, the operator has to defines the region corresponding to the kidney by manually selecting the appropriate radiodensity range. Points with radiodensity in the selected range will be considered as belonging to the kidney. In Materialize Mimics InPrint, this procedure can be performed using the ``\textit{Threshold}" functionality and setting an appropriate HU-based threshold. HU stands for the ``\textit{Hounsfield Unit}" which is the scale used to quantitatively describe radiodensity in an image. This number, computed from the absorption coefficient of the material under standard conditions, enables us to distinguish between different tissues and structures, based on their composition.
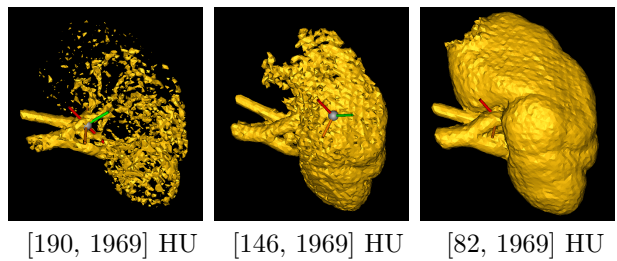
\begin{figure}[t]
\centering
\begin{minipage}{.157\textwidth}
  \centering
  \includegraphics[width=1\linewidth]{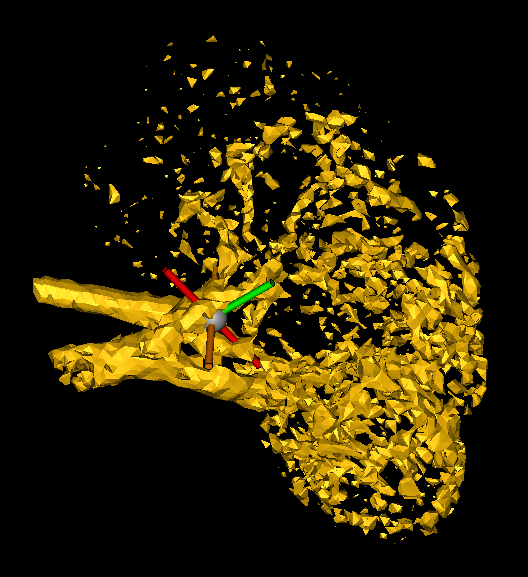}
  \begin{picture}(0,0)
    \put(-30,0){\textcolor{black}{[190, 1969] HU}}
  \end{picture}
\end{minipage}
\hfill
\begin{minipage}{.157\textwidth}
  \centering
  \includegraphics[width=1\linewidth]{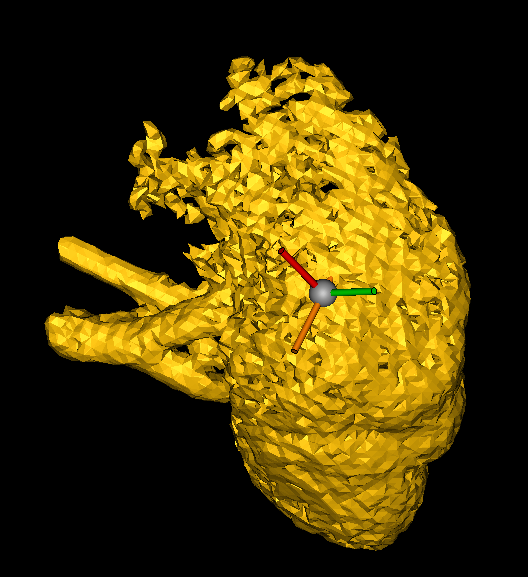}
  \begin{picture}(0,0)
    \put(-30,0){\textcolor{black}{[146, 1969] HU}}
  \end{picture}
\end{minipage}
\hfill
\begin{minipage}{.157\textwidth}
  \centering
  \includegraphics[width=1\linewidth]{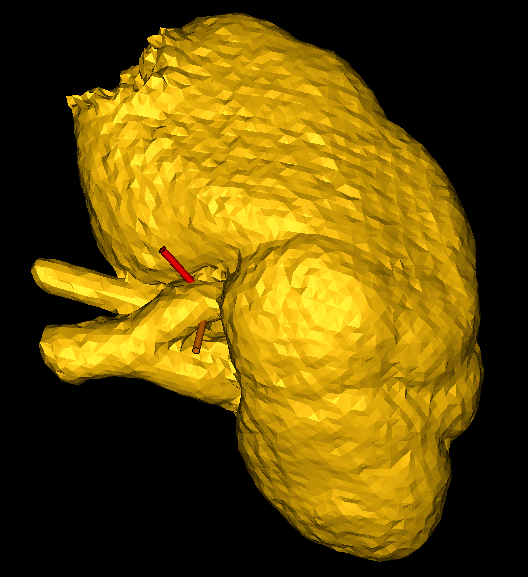}
  \begin{picture}(0,0)
    \put(-27,0){\textcolor{black}{[82, 1969] HU}}
  \end{picture}
\end{minipage}
\caption{It can be observed that as the lower bound varies, the accuracy of the reconstruction differs. In particular, moving from left to right, there is a decreasing of the lower bound, resulting in a more defined and precise reconstruction of the kidney.}
\label{fig:thre}
\end{figure}
The operator can set the HU range by choosing an upper and a lower bound to highlight all the structures of the kidney and minimize cavities and holes in the kidney model (see Figure \ref{fig:thre}).
Additionally, after setting these values, it is advisable to select an option to retain only large regions and automatically fill small holes.
Once the segmentation of the region of interest (bounding box) is performed, the result is the 3D model of the kidney together with other smaller parts from the surrounding tissues (see Figure \ref{fig:kidneysegmentation}).
Any non-kidney components can be easily manually removed. This can be achieved by using a specific command for deletion, allowing these parts to be removed directly from the 3D reconstruction or from the 2D images. In the first case, by selecting the part to be removed, while in the second case, by deleting it slice by slice until it is completely eliminated.\\

\begin{figure}[t]
    \centering
    \includegraphics[width=\columnwidth]{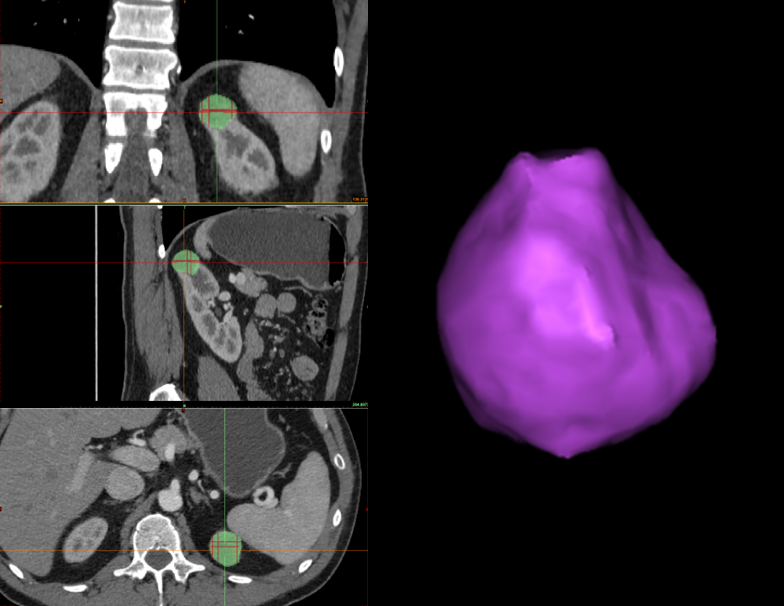}
    \caption{On the left is the automated 3D interpolation of the tumor, based on the manually drawn silhouette in the three planes. On the right is the 3D reconstruction of the tumor.}
    \label{fig:tumorsegmentation}
\end{figure}

\textit{Tumor Segmentation}
\\
In this step, similar to kidney segmentation, a combination of semi-automatic work performed with the tool and manual work carried out by the operator is involved. However, a different procedure is employed for tumor segmentation compared to kidney segmentation.
In this case, the tumor contour is directly outlined. Initially, the operator must identify the tumor in all three sections (axial, coronal, and sagittal) from the CT scan. Subsequently, he should outline the tumor boundary on each of these sections using the ``3D interpolate'' tool available in Materialize Mimics Inprint.
Specifically, when delineating the tumour boundary, Materialise implemented software that semi-automatically segments the tumour over the various slices and produces a 3D reconstruction of the tumour that may change depending on the selection of the tumour boundary in the next slice (see Figure \ref{fig:tumorsegmentation}).
\begin{figure}[t]
     \centering
     \begin{subfigure}{0.23\textwidth}
         \centering
         \includegraphics[width=\textwidth]{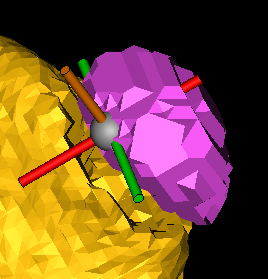}
     \end{subfigure}
     \hfill
     \begin{subfigure}{0.23\textwidth}
         \centering
         \includegraphics[width=\textwidth]{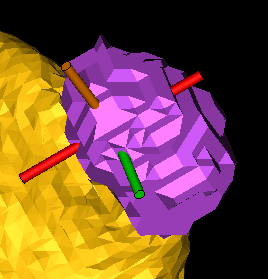}
     \end{subfigure}
        \caption{Here is an example highlighting the importance of outlining the correct perimeter of the tumor. In the left image, the reconstructed tumor exhibits a missing part, leading to a gap between the tumor and the kidney. The right image depicts the tumor after the missing part has been manually added.}
        \label{fig:tum}
\end{figure}
After the automatic reconstruction, the operator should manually inspect the model. If any sections are missing or excessive, they should be corrected manually (see Figure \ref{fig:tum}).\\

\textit{Reconstructions Refinement}
\\Since the tumor and organ were reconstructed independently, it is essential to check for inconsistencies. These could involve irregularities in the kidney's boundary, holes on the kidney interface with the tumor, overlapping, or a lack of contact between the two.
\begin{figure}[t]
    \centering
    \includegraphics[width=\columnwidth]{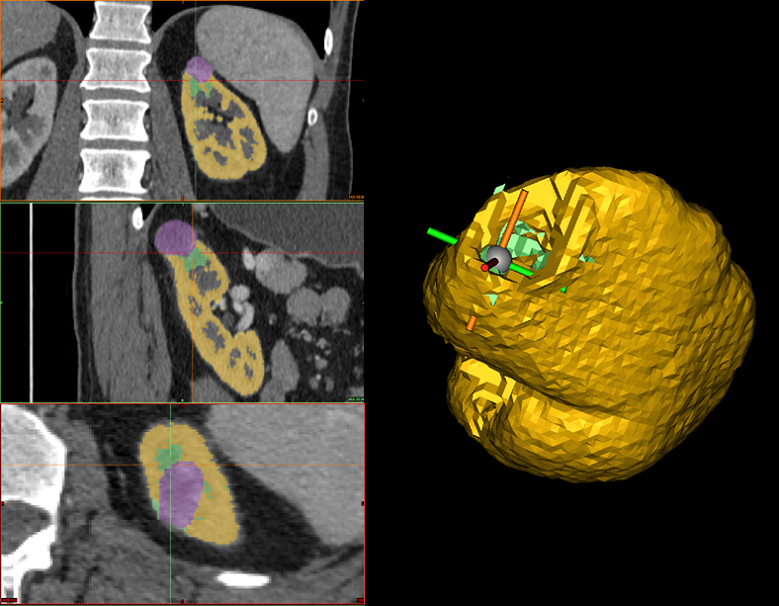}
    \caption{On the left is re the holes filling process across the three different planes. On the right is the 3D reconstruction of the kidney after the filling process.}
    \label{fig:fillingkidney}
\end{figure}
Therefore, the first step in correcting the reconstruction is to remove any tumor parts mistakenly considered part of the organ during kidney segmentation. Then, the operator should inspect the interface between the tumor and the kidney to confirm the absence of gaps. If any holes are present, the ``Fill" command is used to address them while checking each slice across the three sections (see Figure \ref{fig:fillingkidney}). Finally, before exporting the 3D models, a smoothing operator is applied to slightly refine the geometries of the two components. 
The final result can be seen in Figure \ref{fig:finalreconstruction}, which is obtained from a combination of automated and manual segmentation of the organ and the tumor.\\
As already introduced, the organ and tumor segmentation is fundamental for the algorithm that we will present in Section \ref{sec:Algorithm}. Here we have described a manual procedure for this segmentation, however, in the future this could be substituted by a completely automated one.

\section{Algorithm}
\label{sec:Algorithm}
The approach presented in this Section can estimate the CSA between an organ and a tumor using their 3D models, which were reconstructed as described in the previous section. 
For our approach to be effective, these models should be accurate, non-hollow, and maintain their relative positions. 
Although our approach is introduced here for computing the CSA between an organ and a tumor, the algorithm can be applied to any type of 3D object as long as they satisfy the previously listed requirements.
In 3D computer graphics, the shape of an object is defined by a polygonal mesh (PM), typically composed of triangular faces. The number of faces (N) depends on the complexity of the mesh: 
\begin{equation} 
PM = \{F_{1}\dots F_{j} \dots F_{N}\}. 
\end{equation}
A face, according to its shape, is characterized by a set of M vertices. 
\begin{equation}
    F_j = \{V_{j1} \dots V_{jk} \dots V_{jM}\}.
\end{equation}
Where each vertex is defined in the 3D Cartesian space: 
\begin{equation}
    V_{jk} = \{V_{jk}.x, V_{jk}.y, V_{jk}.z\}.
\end{equation}
 
Our approach for computing the CSA between two objects, $O_1$ and $O_2$, begins with this concept. We assume that each object is described by a $PM_i$, which is composed of $N_i$ planar faces (with $i \in [1,2]$)
\begin{equation}
    PM_i = \{F_{i1}, \dots F_{ij}, \dots F_{iN_i}\}.
\end{equation}
Each face $F_{ij}$ is described by a ordered set vertices of dimension $M_{ij}$
\begin{equation}
    F_{ij} = \{V_{ij1} \dots V_{ijk} \dots V_{ijM_{ij}}\},
\end{equation}
and each vertex is expressed in Cartesian coordinates.

The CSA definition requires six intermediate steps, with the first three steps involving the definition and computation of intermediate elements:
\begin{enumerate}
    \item \textit{Centroids computation}: the centroids of each mesh composing the object are computed. This  is done both for the kidney and the tumor.
    \item \textit{Centroid to centroid distance computation}: for each centroid of the smaller object we calculate the distance with the closest centroid from the other object. 
    \item \textit{Threshold computation}: we compute the threshold, defining centroind belonging to the CSA, by analyzing the centroid to centroid distance distribution.
\end{enumerate}
While the last three steps involve the definition, correction, and computation of the CSA:
\begin{enumerate}[resume]
    \item \textit{CSA definition}: we identify the distances computed in Step 2 that are smaller than the threshold value to obtain the list of faces belonging to the CSA. 
    \item \textit{CSA refinement}: we inspect if there are disconnected meshes in the CSA, and in such cases, we determine if they belong to the CSA or not.
    \item \textit{CSA computation}: we calculate the CSA by summing up the area of all the faces that compose it. 
\end{enumerate}

\subsection{Centroids}
Our method works at the face level. However, since we do not make any assumptions about the number of vertices composing the face, we need to extract a coherent descriptor for each face whose structure is independent of the number of vertices. Therefore, we selected the face centroid $C_{ij}$ as the descriptor, and we computed it for each face of both meshes using the following procedure:\\

\begin{algorithmic}[1]
\For{ $\textbf{each}$ $PM_i$ $\in$ \{$PM_1$,$PM_2$\}}
\For{j=1 to $N_i$}
\State $x \gets 0$
\State $y \gets 0$
\State $z \gets 0$
\For{k=1 to $M_{ij}$}
\State $x \gets x + V_{ijk}.x$
\State $y \gets y + V_{ijk}.y$
\State $z \gets z + V_{ijk}.z$
\EndFor
\State $C_{ij}.x \gets x/M_{ij}$
\State $C_{ij}.y \gets y/M_{ij}$
\State $C_{ij}.z \gets z/M_{ij}$
\EndFor
\EndFor
\end{algorithmic}

In this step, the 3D coordinates (x,y and z) of the centroid are calculated. Each coordinate is determined as the sum of the corresponding coordinate of each vertex (lines 6-10) composing the considered face, divided by the total number of vertices (lines 11-13).

\subsection{Centroid-to-Centroid distance}
\label{distance}
To determine which faces belong to the CSA, we use the distance between faces of the two objects.
At this stage, we compute the distance between each face in the mesh of the smaller object (usually the tumor) and the closest face belonging to the other object.  
The distance between two faces is computed as the Euclidean distance between the two faces' centroids.
The result of this procedure is saved in a vector $D = {d_1, \dots, d_L}$, where $L = min(N_1, N_2)$.\\

\begin{algorithmic}[1]
\State $p \gets argmin(N_1,N_2)$
\State $q \gets argmax(N_1,N_2)$
\For{i=1 to $N_p$}
\For{j=1 to $N_q$}
\State $Dt_j \gets euclidean\_distance(C_{pi},C_{qj})$
\EndFor
\State $D_i \gets min(Dt)$
\EndFor
\end{algorithmic}

In lines 1-2, we determine which of the two PM has the maximum number of faces and which has the minimum. In lines 3-6, we calculate the distances between each face of the smaller PM and all the faces comprising the other PM. Then, in line 7, we select only the minimum distance. This results in a vector containing, for each face, only the minimum distances.

\subsection{Threshold}

\begin{figure}[t]
    \centering
    \includegraphics[width=\columnwidth]{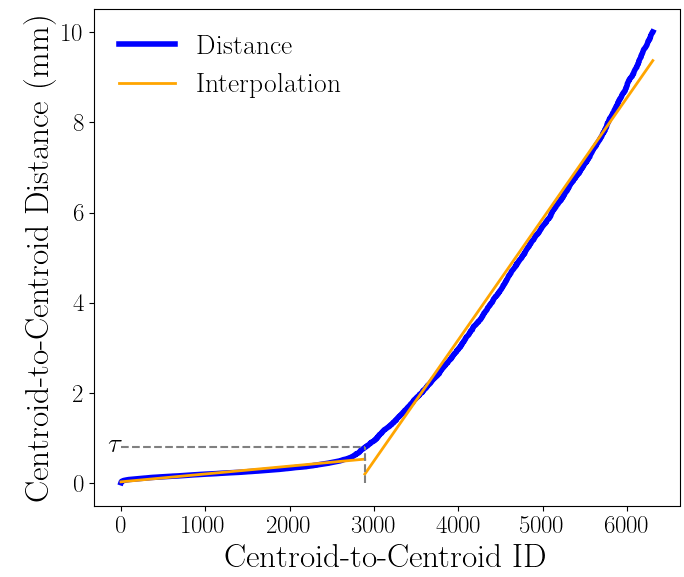}
    \caption{Here is a visual representation of the threshold finding process results. The blue line shows the distribution of centroid-to-centroid distances, while the orange lines represent the closest approximation to the distance distribution.}
    \label{fig:threshold}
\end{figure}

As previously stated, our approach aims to identify which faces from the smaller mesh belong to the CSA. As we will see in Section \ref{definition}, this is achieved thanks to a threshold $\tau$ applied on the centroid-to-centroid distance computed in Section \ref{distance}. However, due to variations in the distance distribution between different object pairs and the potential for reconstruction imperfections, a static threshold is not appropriate. We have devised a method to calculate the threshold for each object pairs by ordering the distance vector, $D$, and analyzing the distance distribution. Figure \ref{fig:threshold} provides a visual representation of this process.

As shown in the figure, the sorted distances begin low and then increase with a clear discontinuity. This discontinuity results from the transition between faces that belong to the CSA and those that do not. We determine the threshold value by identifying the two lines that best approximate the distance distribution and finding their intersection point. This is achived by dividing the sorted distances into two parts and performing linear interpolation on both parts. For the linear interpolation, we adopt a least square polynomial approach (\textit{lsq\_fit}). We vary the division point over all the samples in the distance vector to find the best division point and resulting threshold. To increase precision and reduce the search space, we conduct the threshold search on a subset of the sorted distances, those less than 1 cm.\\

\begin{algorithmic}[1]
\State $Dt \gets quicksort(D)$
\State $F \gets 0$
\For{i=1 to L}
\If{$Dt_i < 1cm$}
\State $Ds_i \gets Dt_i$
\State $F \gets F+1$
\Else
\State break
\EndIf
\EndFor

\For{i=2 to F-1}
\State $f1 \gets lsq\_fit(Ds_{[1,i]})$
\State $f2 \gets lsq\_fit(Ds_{[i+1,F]})$
\State $\overline{D}_{[1,i]} \gets f1(1,i)$
\State $\overline{D}_{[i+1,F]} \gets f2(i+1,F)$
\State $d_i \gets 0$
\For{j=1 to F}
\State $d_i \gets d_i + euclidean\_distance(Ds_j,\overline{D}_j)$
\EndFor
\EndFor
\State $id = argmin(d)$
\State $\tau = Ds_{id}$
\end{algorithmic}

In lines 3-5, we select the distance values less than 1 cm. In lines 11-15, we perform the interpolation using ``lsq\_fit". Then, in lines 16-19, we compute the point-to-point distances between the actual distance values and the interpolated ones, measuring the cumulative error for each case. In line 21, we choose the id with the minimum cumulative error as the one identifying our threshold.
\subsection{CSA definition}
\label{definition}
Once the threshold $\tau$ has been identified, it is possible to determine which faces from the smaller mesh are part of the CSA. This is simply done by selecting only those faces associated with a centroid-to-centroid distance lower than the identified threshold. The outcome of this phase is a list of the IDs of the faces that belong to the CSA.\\

\begin{algorithmic}[1]
\State $IDs \gets \{\}$
\State $j \gets 1$
\For{i=1 to L}
    \If{$D_i < \tau$}
    \State $IDs_j \gets i$
    \State $j \gets j+1$
    \EndIf
\EndFor
\end{algorithmic}

In line 4, we check if the distances computed in Subsection \ref{distance} are less than the threshold value. When this condition is met, the ID is inserted in the list created in line 1 (line 5). The output is a list containing all the IDs of faces belonging to the CSA.

\subsection{CSA refinment}
\begin{figure}[t]
\centering
\begin{subfigure}[b]{0.23\textwidth}
    \includegraphics[width=\textwidth]{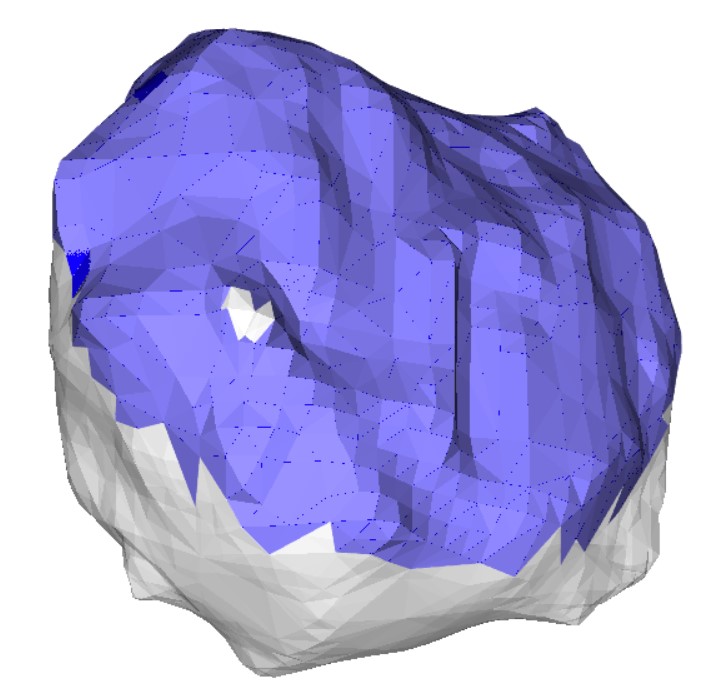}    
    \caption{pre refinment}
    \label{fig:ref_beg}
\end{subfigure}
\hfill
\begin{subfigure}[b]{0.23\textwidth}
    \includegraphics[width=\textwidth]{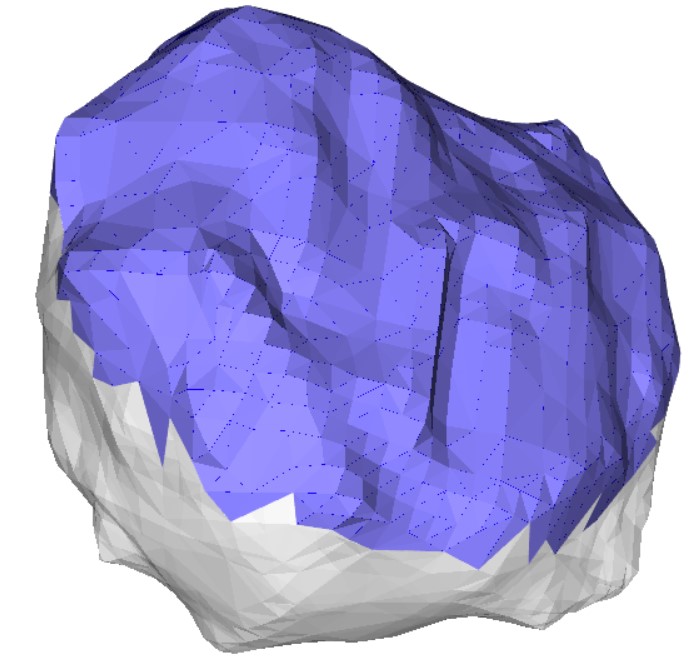}    
    \caption{post refinement}
    \label{fig:ref_end}
\end{subfigure}
\caption{On the left, the CSA is displayed over a tumor before refinement. It can be noticed that the CSA definition divided the original mesh into three surfaces. On the right, the CSA after refinement shows that one of the two surfaces, which was considered non-contact at the beginning, has been associated back to the CSA.}
\label{fig:ref}
    
\end{figure}

The CSA definition divides the faces of the smaller mesh into two sets: points in contact with the other mesh and points not in contact. The two sets should consist of faces connected to each other due to the nature of the problem. However, imprecision in the reconstruction can sometimes cause the CSA definition to divide the mesh into more than two connected areas, as shown in Figure \ref{fig:ref_beg}. Therefore, our approach always checks if there are more than two fully connected areas and refines the CSA to resolve this problem if necessary.

To achieve this, the mesh obtained by removing the CSA faces is inspected. If the resulting mesh is fully connected, nothing is done. However, if the resulting mesh is composed of a set of disconnected meshes, the different meshes are processed to determine whether they should belong to the CSA or not.

First, all the vertices of the faces that are not part of the CSA are added into a graph $G$ (line 4). It is checked if the graph is fully connected. If the graph is fully connected nothing is done, otherwise connected sub-graphs are identified and the corresponding face ids are extracted (line 6).\\

\begin{algorithmic}[1]
\For{i=1 to L}
    \If{$i \notin IDs$}
    \For{k=1 to $M_{pi}$-1}
        \State \textit{G.add\_edge}($V_{pik}$,$V_{pi(k+1)}$)
    \EndFor
    \State \textit{G.add\_edge}($V_{piM_{pi}}$,$V_{pi0}$)
    \EndIf
\EndFor
\If{\textit{not} \textit{isconnected(G)}}
\State $Sm \gets \textit{connected\_components}(G)$
\EndIf
\end{algorithmic}

This procedure identifies a list of sub-meshes named $Sm$ (line 10). The length of the list is equivalent to the number of sub-meshes identified by the \textit{connected\_components} method. Each element within the list is also a list comprising all the IDs of the faces that make up that specific sub-mesh. 

At this point, we need to determine which sub-meshes should be considered part of the CSA. Since our method should divide the original mesh into two sub-meshes: the surface in contact with the other model and the surface not in contact. If, after removing the CSA, we have more than one sub-mesh (i.e., $|Sm| > 1$), it means that the remaining sub-meshes contain points that the thresholding mechanism wrongly labeled as not belonging to the CSA. Therefore, for each sub-mesh, we look for the faces with the highest centroid-to-centroid distance $D_i$, and we pick the mesh with the furthest face as the one representing the non-contact surface. All the other sub-meshes' faces are added to the CSA, as shown in Figure \ref{fig:ref_end}.\\

\begin{algorithmic}[1]
\State $t \gets \{0\}$
\For{i=1 to $|Sm|$}
    \For{j=1 to $|Sm_i|$}
        \If{$t_i \leq D_{Sm_{ij}}$}
            \State $t_i \gets D_{Sm_{ij}}$ 
        \EndIf
    \EndFor
\EndFor

\State $not\_csa = argmax(t)$
\For{i=1 to $|Sm|$}
    \If{$j \neq not\_csa$}
    \For{j=1 to $|Sm_i|$}
        \State $k \gets length(IDs)$
        \State $IDs_{k+1} \gets j$
    \EndFor
    \EndIf
\EndFor
\end{algorithmic}

In lines 2-8, we select for each sub-mesh the face with the highest distance to the CSA, and in line 9, we designate the one with the maximum distance as the surface not in contact. In lines 10-17, we add the faces id from the other sub-meshes to the list of faces IDs belonging to the CSA.

\subsection{CSA computation}
After ensuring that the ids of all the faces belonging to the CSA are stored in the $IDs$ vector, the overall area of the CSA is computed. To do this, we sum up all the face areas, as shown below (lines 2-5):\\

\begin{algorithmic}[1]
\State $CSA \gets 0$
\For{i=1 to $|IDs|$}
    \State $j \gets IDs_i$
    \State $CSA \gets CSA + area(C_{pj})$
\EndFor
\end{algorithmic}

Since we aimed to preserve generality, we made no assumptions about the shape of mesh faces. Although most meshes have triangular faces, we decided to be agnostic to this fact. To compute the area of a single face, we convert the vertices from 3D to 2D by projecting them onto the face plane. Then, we use the Shoelace formula to compute the area as follows:
\begin{equation}
    A = \frac{1}{2}|\sum_{i=1}^{n-1}x_iy_{i+1} + x_ny_1 - \sum_{i=1}^{n-1}x_{i+1}y_i - x_1y_n|
\end{equation}
where A represents the area of the polygon, n represents the number of vertices of the polygon, and x and y are the coordinates of the vertices in the face plane.

\section{Evaluation}
\label{sec:Evaluation}
The solution described in the previous section has been developed into a Python class and integrated into a graphical user interface for ease of use. Due to the limitations of the libraries used to manage the STL files, the current implementation only supports triangular meshes. However, the code is written in a way that preserves the solution's nature and allows for future extensions to non-triangular meshes. The GUI provides a simple way to use our approach. It allows users to load the mesh of the tumor and the organ, computes the CSA, provides a 3D visualization of the CSA over the two original models, and returns additional useful statistics such as the overall tumor area and volume. The source code for the GUI is publicly available on GitHub\footnote{\url{github.com/ACarfi/contact-surface-area-gui}}. From the GitHub repository, it is also possible to download an executable, which makes it easier for practitioners to adopt the approach.

\begin{figure}[t]
    \centering
    \includegraphics[width=\columnwidth]{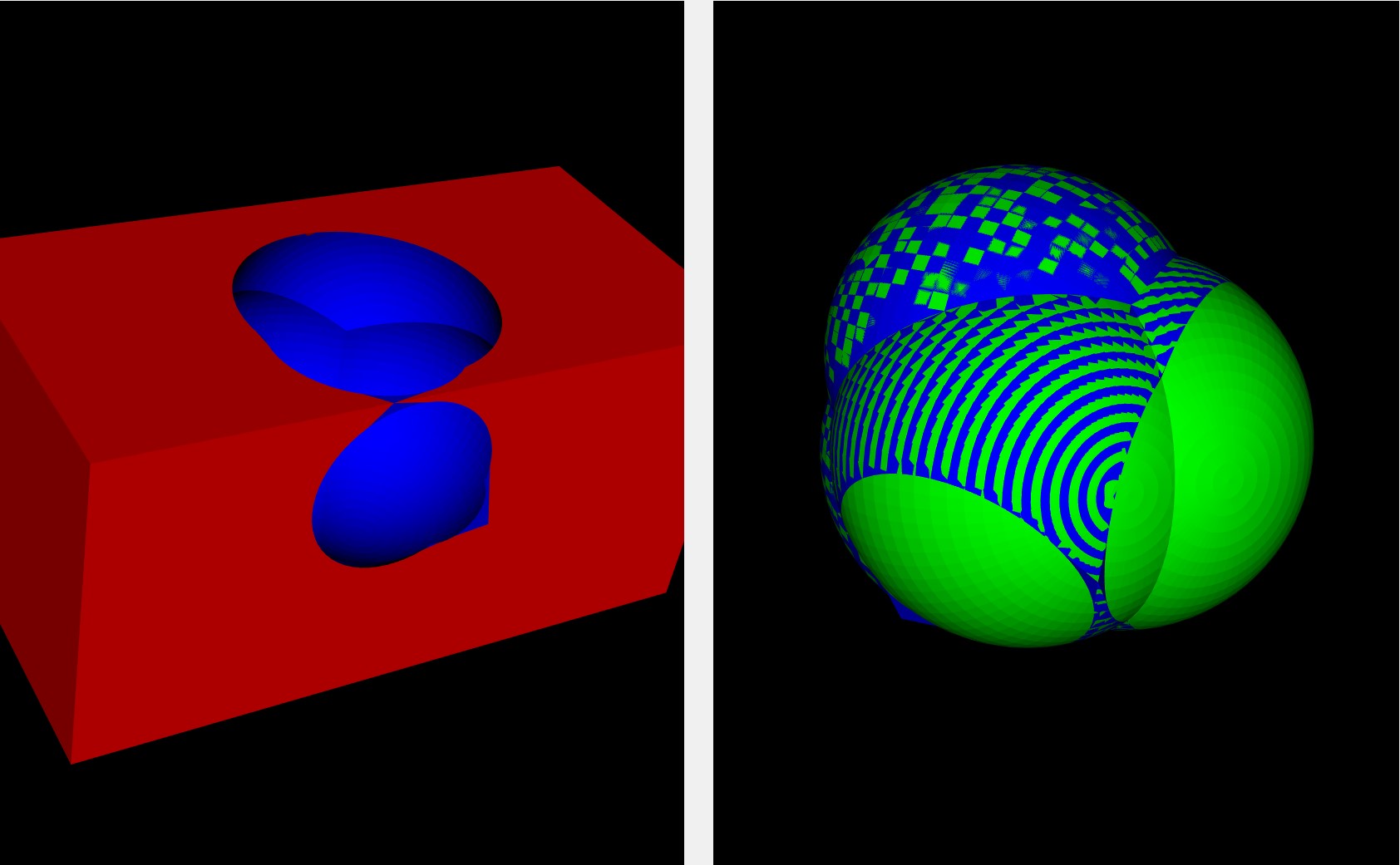}
    \caption{A screenshot of our graphical user interface displays an example of the synthetic benchmark and the resulting CSA analysis.}
    \label{fig:benchmark}
\end{figure}

\begin{figure}[t]
    \centering
    \includegraphics[width=\columnwidth]{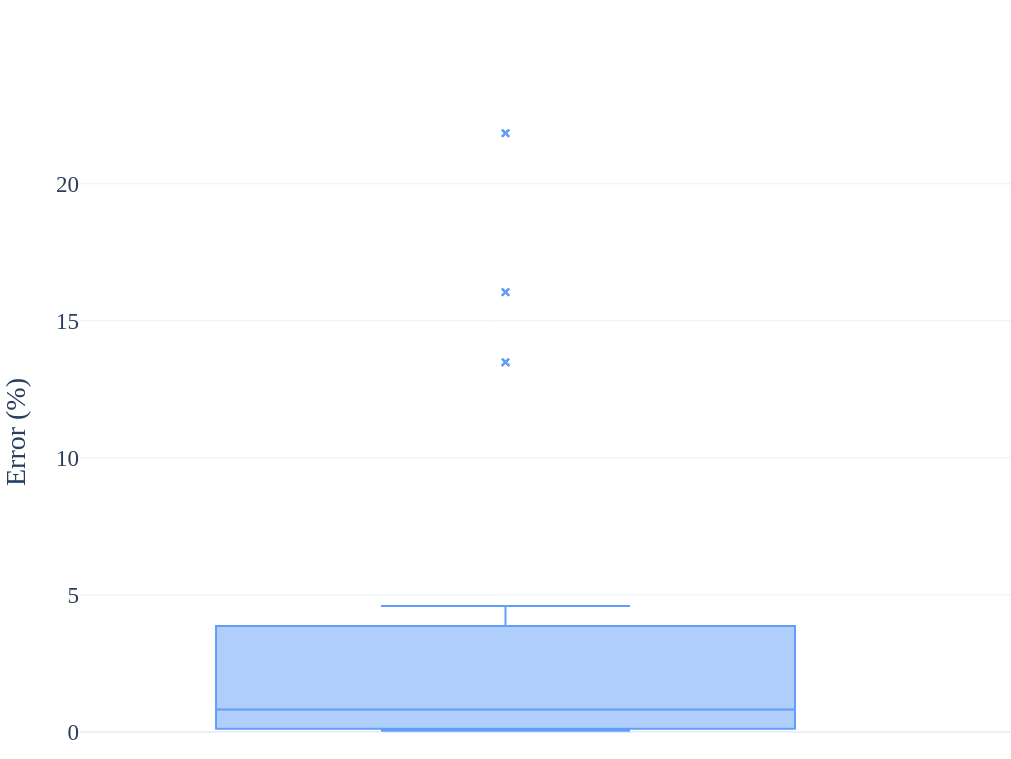}
    \caption{The box plot shows the percentage errors observed in the synthetic benchmark, with three outliers marked with a cross.}
    \label{fig:error}
\end{figure}

To evaluate the accuracy of our approach, we conducted two types of evaluation: objective and subjective. In the objective evaluation, we tested the precision of the CSA computation using a set of synthetic models. These models were created using CAD software, which allowed us to know the contact surface area. The synthetic benchmark consists of twenty pairs of models and is publicly available on the previously referenced GitHub repository. The synthetic organ is always a rectangular base prim with a cut corresponding to the synthetic tumor, while the synthetic tumor varies from a simple sphere to more complex shapes. Figure \ref{fig:benchmark} shows a screenshot from our GUI that displays an example of synthetic models.

The results of the objective evaluation are reported in the box plot of Figure \ref{fig:error}. As can be seen from the box plot, with the exception of a few outliers, the system is characterized by low percentage errors, with a median percentage error close to zero.

The subjective evaluation was conducted to assess the overall procedure, which includes tumor and organ reconstruction, along with contact surface area computation. However, it is not possible to obtain CSA ground truths for a real couple of tumor and organ. Therefore, the precision of CSA was subjectively evaluated by a trained operator. We reconstructed 87 couples of organ-tumor from 82 patients undergoing partial kidney nephrectomy and fed the resulting 3D models to our GUI. Upon visual inspection, the results for each of the 87 couples were qualitatively acceptable.

In order to conduct the subjective evaluation of the algorithm, the study was submitted for approval to the ethics committee of Policlinico San Martino (Ethics Committee code: PT44; regional number: 554/2023).

\section{Conclusions} 
\label{sec:Conclusion}
The approach proposed in this study provides a precise formalization of the CSA computation when accurate 3D models of the organ and tumor are available. The main novelty of our approach is that, contrary to previous solutions, it does not require assumptions about the tumor's shape. Since the algorithm for computing the CSA uses the geometries of both the tumor and organ, it is fundamental for the reconstruction to be as accurate as possible. In our article, we present a protocol to follow, which we executed using the software Materialize. Alternative medical 3D reconstruction software could also be adopted. Although our work has been designed and tested for kidney tumors, we want to point out that the algorithm described here is general and could be applied to other organs as well.

Compared to other pre-existing solutions, our approach aims to reduce the influence of human error in CSA computation. State-of-the-art solutions rely on values that must be manually measured by a human operator, which can decrease both the accuracy of the CSA estimate and the reproducibility of studies. Even methods based on 3D reconstruction require manual annotation of the model to define the CSA, making it difficult to reproduce calculations. Although our approach still requires human labor for creating 3D reconstructions, human intervention is only necessary for that step, not for defining the CSA. Moreover, researchers are already working on the automation of organ and tumor segmentation from CT scans for 3D reconstruction, and a few commercial products are available for specific organs. As a result, in the near future, when 3D reconstructions can be performed autonomously, our approach will enable the computation of the CSA without any human intervention.

In this article, we present both a quantitative and a qualitative analysis of our algorithm. Using a synthetic dataset created by us, we were able to show a low percentage of errors in our algorithm. Additionally, by visually inspecting the results on 87 pairs of organ-tumor, we were able to determine if the algorithm behaves properly on real-world data. It is important to note that the synthetic data we used for our evaluation is publicly available and can be used in future studies to evaluate new CSA computation approaches. To ensure maximum reusability of our results, we have made the implementation of our code available as open source. Finally, we also provide a graphical user interface that makes it easy to compute the CSA by providing two STL files.

\section{Acknowledgment}
This research was made, in part, with the Italian government support under the National Recovery and Resilience Plan (NRRP), Mission 4, Component 2 Investment 1.5, funded from the European Union NextGenerationEU.

 \bibliographystyle{elsarticle-num}

\end{document}